\documentclass[twocolumn,preprintnumbers,amsmath,amssymb,prl,longbibliography, superscriptaddress]{revtex4-1}
\usepackage{amssymb}
\usepackage{graphicx}
\usepackage{bm}
\usepackage{braket}
\usepackage{color}
\usepackage[breaklinks=true,colorlinks=true]{hyperref}

\makeatletter
  \def\Tr{\mathop{\operator@font tr}\nolimits}
\makeatother
\def\trB{\mathop{\Tr_\mathrm{B}}}
\def\trA{\mathop{\Tr_\mathrm{A}}}

\newcommand{\mrA}{\mathrm{A}}
\newcommand{\mrB}{\mathrm{B}}
\newcommand{\mrc}{\mathrm{c}}

\begin{document}
\preprint{\texttt{IPMU17-0027}}

\title{\textbf{\textsf{Universality in volume-law entanglement of scrambled pure quantum states}}}
\author{Yuya O. Nakagawa}
\email{y-nakagawa@issp.u-tokyo.ac.jp}
\affiliation{Institute for Solid State Physics, The University of Tokyo. Kashiwa, Chiba 277-8581, Japan.}
\affiliation{Department of Physics, Faculty of Science, The University of Tokyo. Bunkyo-ku, Tokyo 133-0022, Japan.}
\author{Masataka Watanabe}
\affiliation{Kavli Institute for the Physics and Mathematics of the Universe (WPI),The University of Tokyo Institutes for Advanced Study, The University of Tokyo, Kashiwa, Chiba 277-8583, Japan}
\affiliation{Department of Physics, Faculty of Science, The University of Tokyo. Bunkyo-ku, Tokyo 133-0022, Japan.}
%
\author{Hiroyuki Fujita}
\affiliation{Institute for Solid State Physics, The University of Tokyo. Kashiwa, Chiba 277-8581, Japan.}
\affiliation{Department of Physics, Faculty of Science, The University of Tokyo. Bunkyo-ku, Tokyo 133-0022, Japan.}
\author{Sho Sugiura}
\email{sugiura@issp.u-tokyo.ac.jp}
\affiliation{Institute for Solid State Physics, The University of Tokyo. Kashiwa, Chiba 277-8581, Japan.}

\maketitle

\section*{abstract}
A pure quantum state can fully describe thermal equilibrium as long as one focuses on local observables. The thermodynamic entropy can also be recovered as the entanglement entropy of small subsystems. When the size of the subsystem increases, however, quantum correlations break the correspondence and mandate a correction to this simple volume law. The elucidation of the size dependence of the entanglement entropy is thus essentially important in linking quantum physics with thermodynamics.
Here we derive an analytic formula of the entanglement entropy for a class of pure states called cTPQ states representing equilibrium. We numerically find that our formula applies universally to any sufficiently scrambled pure state representing thermal equilibrium, i.e., energy eigenstates of non-integrable models and states after quantum quenches. Our formula is exploited as diagnostics for chaotic systems; it can distinguish integrable models from non-integrable models and many-body localization phases from chaotic phases.


\section*{Introduction}
As a measure of the quantum correlations in many-body systems, the entanglement entropy (EE) has become an indispensable tool in modern physics. The EE quantifies the amount of non-local correlation between a subsystem and its compliment. The ability of the EE to expose the non-local features of a system offers a way to characterize topological orders~\cite{KitaevPreskill2006, LevinWen2006}, solve the black hole information paradox~\cite{Hawking_paradox}, and quantify information scrambling in a thermalization process under unitary time evolution~\cite{Hayden2007, Ruihua2016,OTOCexpr}. 

\begin{figure}
\begin{center}
\includegraphics[width=0.9\linewidth]{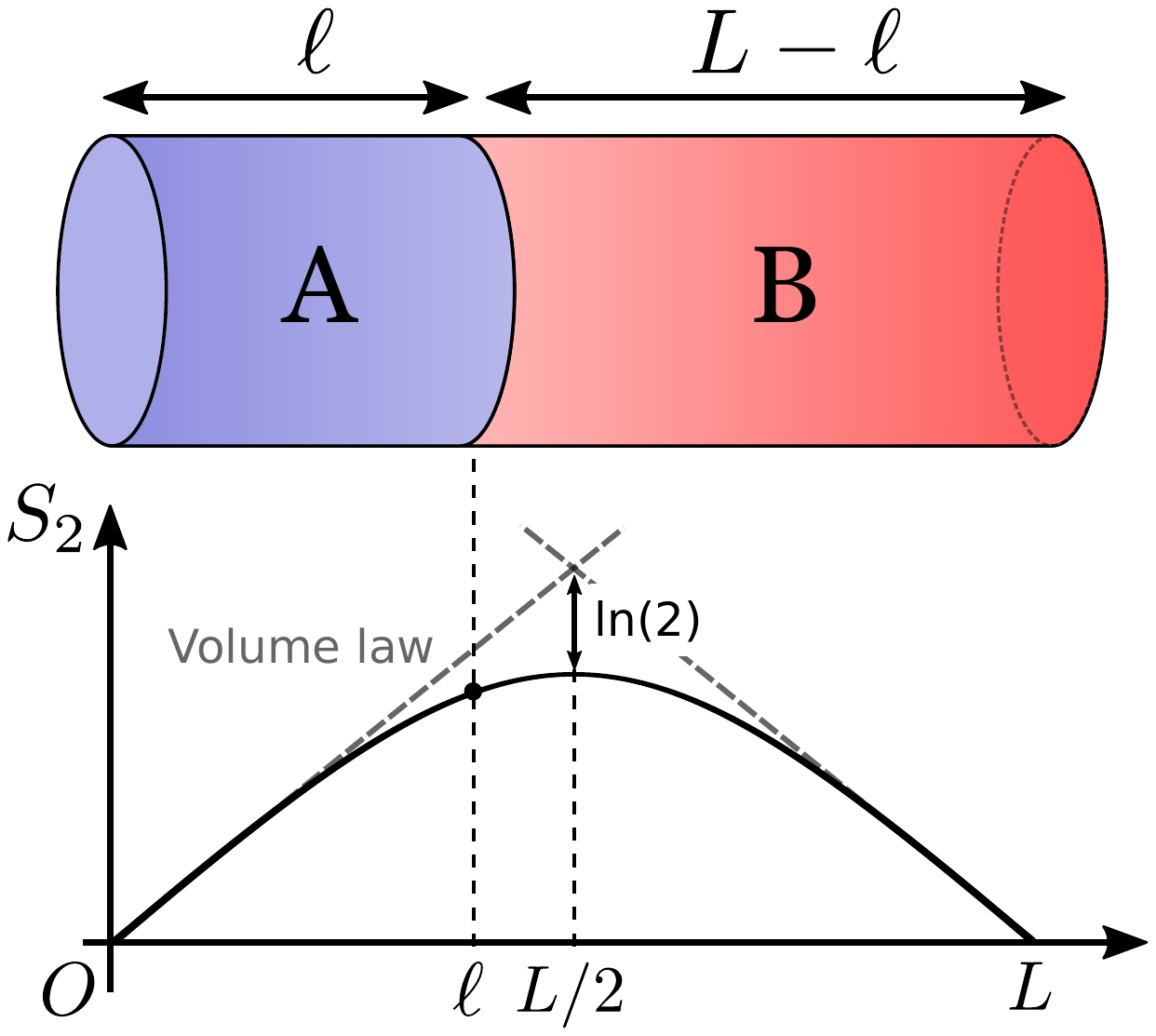}
\end{center}
\vspace{-6mm}
\caption{
{\bf A schematic picture of our setup.}
The second R\'{e}nyi Page curve for pure states, $S_2(\ell)$,
follows the volume law when $\ell$ is small, but gradually 
deviates from it as $\ell$ grows. At the middle, $\ell=L/2$,
the maximal value is obtained, where the deviation from the volume law
is $\ln 2$ (see the Result section). 
Past the middle $\ell=L/2$, it decreases toward $\ell=L$
and becomes symmetric under $\ell \leftrightarrow L - \ell$.
}
\label{PC}
\end{figure}

Recently, the EE was measured in quantum many-body systems for the first time~\cite{MeasurementEE2015, Science2016}. 
Specifically, the second R\'{e}nyi EE (2REE), one of the variants of the EE, of a pure quantum state was measured in quantum quench experiments using ultra-cold atoms.  
For such pure quantum states, it is believed that the EE of any small subsystem increases in proportion to the size of the subsystem just like the thermodynamic entropy~\cite{Eisert2010}. 
This is called the volume law of the EE.  However, when the size of a subsystem becomes comparable to that of its complement, it is observed experimentally that the EE starts deviating from the volume law, and eventually decreases (Fig. \ref{PC}). 
The curved structure of the size dependence of the EE, which first increases linearly and then decreases, universally appears in various excited pure states, for example, energy eigenstates~\cite{Grover2015} and states after quantum quenches~\cite{Calabrese_quench,Takayanagi2010}.
We call this curved structure a Page curve, after D.~Page~\cite{Page:1993df}, 
and it is of fundamental importance in explaining these examples to reveal a universal behavior of the Page curve. 

Despite their ubiquitous appearance, 
the theoretical understanding of the Page curves is limited to the case of a random pure state~\cite{Page:1993df, TypicalEE2014}, which is a state at infinite temperature and can be defined only in a finite-dimensional Hilbert space. 
By contrast, the cold-atom experiments address finite temperature systems
and an infinite-dimensional Hilbert space; therefore, it is important to develop a theory of the Page curve applicable to these situations. 
Additionally, there are practical needs for the estimation of the slope of the volume law. The slope is often employed, for example, to calculate the corresponding thermodynamic entropy~\cite{Science2016,Grover2015} and to detect a transition between the energy eigenstate thermalization hypothesis (ETH) phase and the many-body localized (MBL) phase~\cite{Pollmann2014, MBLETH2016}. However, since the experimentally or numerically accessible sizes of the systems are small, the estimation of the volume-law slope is deteriorated by the curved structure of the Page curve. 

In this work, we show that Page curves in broad classes of excited pure states exhibit universal behaviors. 
We first derive the function of the Page curve for canonical thermal pure quantum (cTPQ) states, which are pure states representing thermal equilibrium at a temperature $\beta^{-1}$~\cite{Sugiura2012,Sugiura2013}. 
In particular, the 2REE is controlled only by two parameters, an effective dimension $\ln a$ and an offset $\ln K$, for any Hamiltonian and at any temperature. 
We conjecture and numerically verify that this feature of the Page curve universally appears in any sufficiently scrambled pure states representing equilibrium states; that is, our function fits the 2REE of the energy eigenstates of a non-integrable system and the states after quantum quenches including the state realized in the above-mentioned experiment \cite{Science2016}. 
By contrast, in the case of the energy eigenstates of an integrable system, which are not scrambled at all, we find that their Page curves deviate from our function. 
Since our function enables us to estimate the slope of the volume law from small systems with high accuracy and precision, 
our result is also numerically effective in detecting the MBL-ETH transition ~\cite{Pollmann2014, MBLETH2016,MBL2015, MBLPalHuse2010}
and improves the estimation of the critical exponent.

\section*{Results}
\subsection*{Derivation of the Page curve in cTPQ states}
Let us consider a lattice $\Sigma$ containing $L\times M$ sites (Fig.~\ref{PC}), equipped with a translation-invariant and local Hamiltonian $H$. 
We divide $\Sigma$ into two parts, 
A and B, each containing $\ell \times M$ and $(L-\ell) \times M$ sites. 
The $n$-th R\'{e}nyi entanglement entropy of a pure quantum state $|\psi \rangle$ is defined as
\begin{align}
S_n(\ell) = \frac{1}{1-n} \ln \left( \Tr \rho_\mrA^n \right),
\end{align}
where $\rho_\mrA \equiv \trB |\psi \rangle \langle \psi |$. 
We call $S_n(\ell)$ as a function of $\ell$ the $n$-th R\'{e}nyi Page curve ($n$RPC). 
We note that 
we use the term differently from
how it is used in the context of quantum gravity, where it denotes the temporal dependence of the entanglement during the formation of a black hole.
To simplify the calculation, we assume $\ell, L \gg 1$.

To derive the behaviors of $n$RPCs for any Hamiltonian, we utilize cTPQ states, 
which are proposed along with studies of typicality in quantum statistical mechanics~\cite{vonNeumannTypicality, Bocchieri1959, Tasaki1998, Goldstein2006, Popescu2006, Sugita, Reimann2007}. 
The cTPQ state at the inverse temperature $\beta$ is defined as
\begin{equation}
	\Ket{\psi}={1\over \sqrt{ Z_\psi}}
	\sum_{j} z_{j}e^{-\beta H/2}\Ket{j},
\label{TPQ def}
\end{equation}
where
$Z_\psi \equiv \sum_{i,j} z_{i}^* z_j \Bra{i} e^{-\beta H} \Ket{j}$ is a normalization constant,
$\{\Ket{j} \}_j$ is an arbitrary complete orthonormal basis of the Hilbert space $\mathcal{H}_\Sigma$, 
and the coefficients $\{z_j\}$ are random complex numbers $z_j \equiv (x_j+i y_j)/\sqrt{2}$, 
with $x_j$ and $y_j$ obeying the standard normal distribution $\mathcal{N}(0,1)$.
For any local observable, the cTPQ states at $\beta$ reproduce their averages in thermal equilibrium at the same inverse temperature~\cite{Sugiura2013}.
As a starting point, we here derive the following exact formula 
of the 2RPC at a temperature $\beta^{-1}$ for any Hamiltonian (see the Methods section for the calculations and results for the $n$RPCs), 
\begin{align}
	S_2(\ell)=
	-\ln\left[
	\frac{\trA{(\trB e^{-\beta H})^2}+\trB{(\trA e^{-\beta H})^2}}{(\Tr{e^{-\beta H})^2}}
	\right]. 
	\label{TPQ_S2}
\end{align}

We also 
give several simplifications of Eq.~(\ref{TPQ_S2}). 
The first step is to decompose the Hamiltonian $H$ as $H=H_\mrA+H_\mrB+H_{\rm int}$,
where $H_\mathrm{A,B}$ are the Hamiltonians of the corresponding subregion and $H_{\rm int}$ describes the
interactions between them. 
Since the range of interaction $H_{\rm int}$ is
much smaller than $\ell$ and $L-\ell$ due to the locality of $H$,
we obtain the simplified expression 
\begin{equation}
	S_2(\ell)=-\ln\left(\frac{Z_\mrA(2\beta)}{Z_\mrA(\beta)^2}+\frac{Z_\mrB(2\beta)}{Z_\mrB(\beta)^2}\right) +\ln R(\beta),
	\label{S2=Z}
\end{equation}
where $R(\beta)$, coming from $H_{\rm int}$, is an $O(1)$ constant dependent only on $\beta$
and $Z_\mathrm{A,B}(\beta) \equiv\Tr_\mathrm{A,B}{(e^{-\beta H_\mathrm{A,B}})}$.

Further simplification occurs through the extensiveness of the free energy,
$-\ln  Z_\mathrm{A,B} /\beta$, which is
approximately valid when $\ell, \, L-\ell \gg 1$.
In the region in question, $\ln Z_\mathrm{A,B}$ is proportional to the volume of 
the corresponding subregion, and thus, we replace $Z_\mrA(2\beta)/Z_\mrA(\beta)^2$
and $Z_\mrB(2\beta)/Z_\mrB(\beta)^2$
with $Q(\beta)a(\beta)^{-\ell}$ and $Q(\beta)a(\beta)^{-(L-\ell)}$, respectively.
Here, $a(\beta)$ and $Q(\beta)$ are $O(1)$ constants dependent only on $\beta$,
and $1<a(\beta)$ 
holds because of the concavity and monotonicity of the free energy.
Finally, we reach a simple and universal expression 
for the 2RPC: 
\begin{equation}  \label{FitFunc}
	S_2(\ell)= \ell\ln a(\beta) -\ln\left(1+a(\beta)^{-L+2\ell}\right) + \ln K(\beta),
\end{equation}
where $K\equiv R/Q$.
This is our first main result.
The same simplifications can be applied to a general $n$RPC (see the Method section).
For example, concerning the 3RPC, we have
\begin{align} 
	S_3(\ell)= \ell {\ln b \over 2}
	-{1 \over 2}\ln\left(1+K_1'{b^{\ell} \over a^{L}}+b^{-L+2\ell}\right)
	+ \ln K_2\rq{},
	\label{S3_simple}
\end{align}
where $b$, $K_1'$ and $K_2\rq{}$ are $O(1)$ constants that depend only on $\beta$.

The significance of Eq.~\eqref{FitFunc} is apparent: 
it tells us that the 2RPC 
is determined by only two parameters: 
$a(\beta)$ and $K(\beta)$. 
The first term 
represents the volume law of entanglement for 
$\ell \leq L/2$, and thus, its slope, $\ln a(\beta)$, is a density of the 2REE 
in the thermodynamic limit $L\to\infty$.
The third term, $\ln K$, represents an offset of the volume law.
The second term  
gives the deviation from the volume law, which stems from the highly non-local quantum correlation between subsystems A and B. Here, we see that the way that quantum correlations appear in cTPQ states is completely characterized by the volume-law slope, $a(\beta)$. As $\ell$ approaches $L/2$, the quantum correction to the volume law becomes stronger and eventually becomes exactly $\ln 2$ at $\ell = L/2$,  independent of the inverse temperature $\beta$ and the Hamiltonian. This is a unique feature of the 2RPC, as we do not observe such universal behaviors in the $n$RPCs for $n\geq 3$. 
With regards to the third term, the offset is the term which comes from the degeneracy of a quantum state at zero temperature. This is referred as the topological entanglement entropy for topological states~\cite{KitaevPreskill2006}. Similarly, $\ln K$ contains the degeneracy term, but it also contains other terms, e.g., $\ln Q$. It is interesting future direction to decode the topological entanglement entropy from $\ln K$. 

In addition, by using Eq.~(\ref{FitFunc}), the mutual information is straightforwardly obtained. Suppose that the state is the cTPQ state Eq.~(\ref{TPQ def}) and the system is divided into three part, A, B and C.
The (second R\'enyi) mutual information between A and B is defined and calculated as
\begin{align}
  I_2 &\equiv S_2^\mrA+S_2^\mrB-S_2^\mathrm{A\cup B} \\
  		&= \ln \left(  {a^{- \ell } + a^{-L+ \ell } \over (a^{-\ell/2} + a^{-L+\ell/2})^2 } \right) + q \ln K, 
\label{MI}
\end{align}
where $S_2^{X}$ is the 2REE in $X$, $q=1$ when $\mrA \cup \mrB$ is connected and $q=0$ when $\mrA \cup \mrB$ is disconnected, $\ell$ is the sum of the length of A and B, and, for simplicity, we take the each length to be $\ell/2$. 
Eq.~(\ref{MI}) explains the observed size-dependence of the mutual information in Ref.~\cite{Science2016}. 
$I_2$ grows exponentially with $\ell$ for $\ell < L/2$, and shows a linear growth for $L/2 < \ell$.  
See Supplementary Note~1 for the detailed explanations. 

\begin{figure}
\begin{center}
\includegraphics[width=\linewidth]{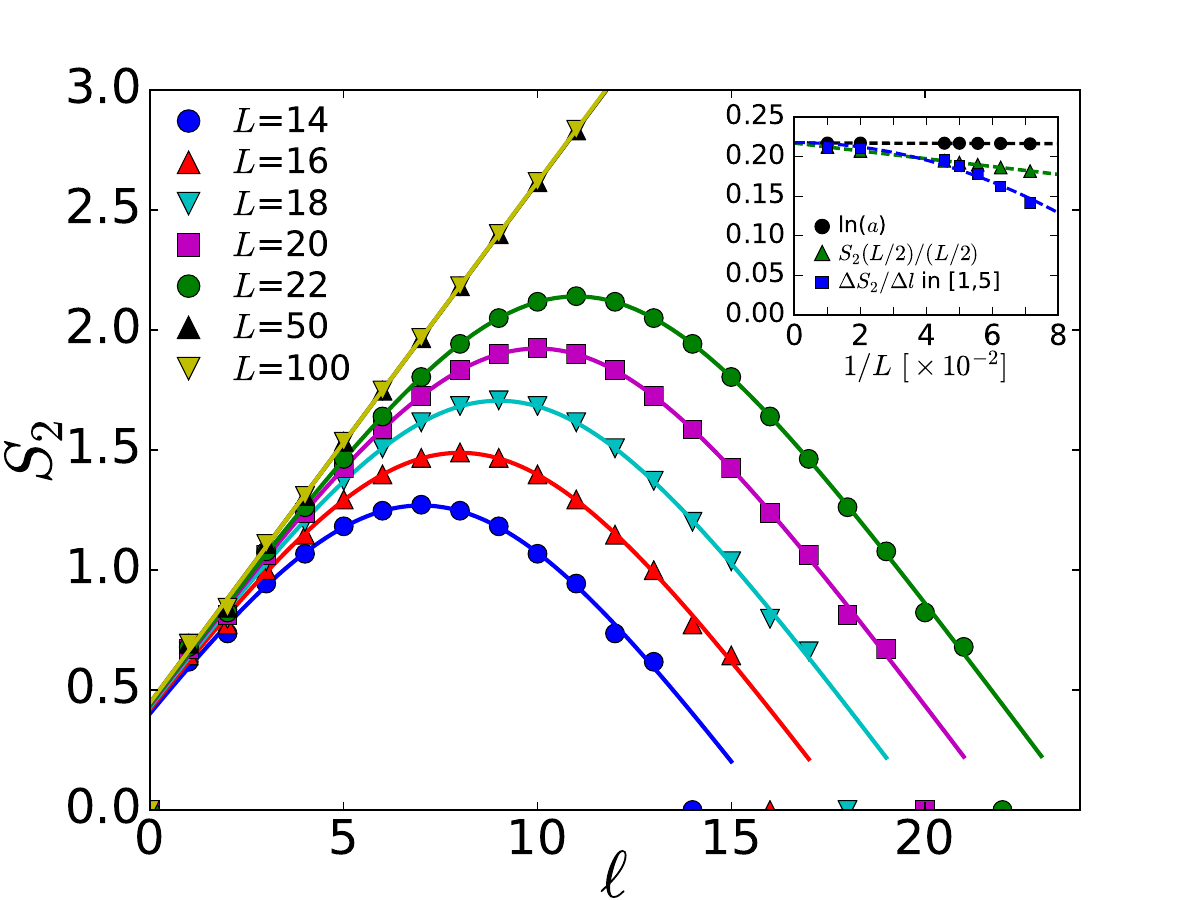}
\end{center}
\vspace{-6mm}
\caption{
{\bf Second R\'{e}nyi Page curve in cTPQ states.}
The dots represent the second R\'{e}nyi Page curves in the cTPQ states of the spin system~\eqref{TPQ_model}
at an inverse temperature $\beta =4$ calculated by Eq.~\eqref{TPQ_S2} for various system sizes $L$.
The lines are the fits by Eq.~\eqref{FitFunc} for the numerical data.
The inset shows the fitted values of $\ln a$, $S_2(L/2)/(L/2)$, 
and the average slope of the curve between $\ell=1$ and $\ell=5$.
The dotted lines are the extrapolations to $L \to \infty$ by $1/L$ scaling for $\ln a$ and $S_2(L/2)/(L/2)$
and by $1/L^2$ scaling for the average slope.
}
\label{Fig TPQ}
\end{figure}

Finally, to confirm the validity of the approximations and clarify the advantages of our formula~\eqref{FitFunc},
we present numerical simulations of the 2RPC
of cTPQ states for the $S=1/2$ XY chain under a periodic boundary condition,
\begin{align}
  H = \sum_{i=1}^L \left(S^x_i S^x_{i+1}+S^y_i S^y_{i+1} \right).
\label{TPQ_model}
\end{align}
This system is mapped to the free fermion system by the Jordan-Wigner transformation~\cite{JWtransformation}, 
and the quantities $\trA(\trB e^{-\beta H})^2$ and $\trB(\trA e^{-\beta H})^2$
can be efficiently calculated in a large system ($L \sim 100$) by the correlation functions of the system~\cite{Peschel2003}.
We numerically calculate the 2RPC of the cTPQ states at the inverse temperature $\beta=4$ by evaluating Eq.~\eqref{TPQ_S2}.
As Fig.~\ref{Fig TPQ} shows, the numerical data of the 2RPC
are well fitted by our formula~\eqref{FitFunc} for all system sizes $L$ and subsystem sizes $\ell$
(details of the fitting is described in the Method section).
In addition,  we compare several estimates of the density of the 2REE
from numerical data in the inset: $\ln a$ from the fits by our formula, 
the density of the 2REE for half of the system, $S_2(L/2)/(L/2)$, 
and the average slope of the curve between $\ell=1$ and $\ell=5$, $(S_2(5)-S_2(1))/4$.
It is clear that $\ln a$ does not contain any systematic error compared with the other two estimates,
which represents one of the advantages of our formula~\eqref{FitFunc}.
We also numerically check the validity of the approximations in deriving the formula~\eqref{FitFunc} in Supplementary Figure~1.

\subsection*{General conjecture for scrambled states}
So far, we have focused on the cTPQ states, but they are merely a canonical example
of pure states locally reproducing the Gibbs ensemble.
Here, we pose a conjecture
for other scrambled pure states: 
Equation~\eqref{FitFunc} 
works as a fitting function for generic scrambled pure states.

In the following two subsections, we numerically check this conjecture. 
We calculate the 2RPCs of various pure states, namely, 
the excited energy eigenstates of (non-)integrable models, and the states after a quantum quench. 
We show that the conjecture holds for the eigenstates of the non-integrable model but not for those of the integrable model. We also numerically reveal that Eq.~\eqref{FitFunc} well fits the 2RPC averaged over the time evolution after a quantum quench. 

\subsection*{Numerical results for energy eigenstates}
As the ETH claims, in a wide class of models, the energy eigenstates look thermal - the expectation values of the local observables reproduce those of the Gibbs ensemble~\cite{ETHDeutch1991, ETHSrednicki1994, Thermalization2008}.
From the viewpoint of ETH, its extension to non-local quantities is interesting~\cite{Grover2015, Scott2016}.
We test whether the formula~\eqref{FitFunc} applies to
the 2RPC, which is highly non-local at $\ell = O(L)$, in particular $\ell \simeq L/2$.

As an example, we take the $S=1/2$ XXZ spin chain with/without next-nearest neighbor interactions
under the periodic boundary condition,
\begin{align}
H = \sum_{i=1}^L \left( S^x_i S^x_{i+1}+S^y_i S^y_{i+1}+  \Delta S^z_i S^z_{i+1} + J_2 \bm{S}_i \cdot \bm{S}_{i+2} \right),
\label{eigenstate_model}
\end{align}
where we set $\Delta =2$ and $J_2=4$ for non-integrable cases and $J_2=0$ for integrable cases~\cite{TakahashiBook}.

\begin{figure*}
   \includegraphics[width = 1.0\linewidth]{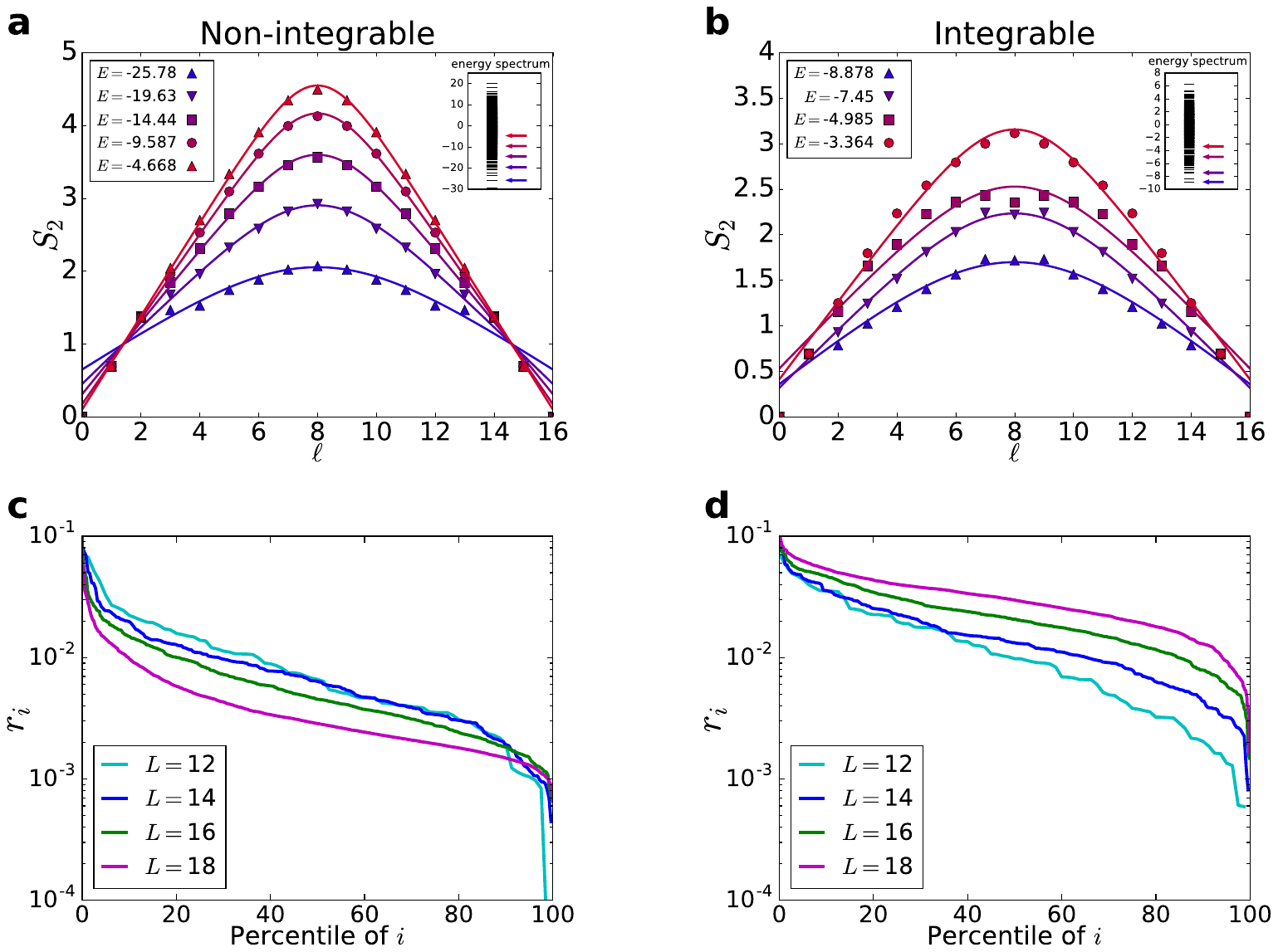}
\caption{
{\bf Second R\'{e}nyi Page curve for general energy eigenstates.}
{\bf a,} 2RPCs of several energy eigenstates of
the non-integrable Hamiltonian, Eq.~\eqref{eigenstate_model} with $\Delta=2$ and $J_2=4$ (dots),
and the fits by our formula \eqref{FitFunc} (lines).
The inset shows the energy spectrum of the Hamiltonian, and the arrows indicate the
eigenstates presented in the figure.
{\bf b,} Same as figure {\bf a} for the integrable Hamiltonian ($\Delta=2, J_2=0$).
{\bf c,} Residuals of fits per site $r_i \equiv L^{-1} \sum_{\ell=0}^L (S_2(\ell)_{i, \mathrm{data}} -  S_2(\ell)_{i, \mathrm{fit}} )^2$, where $S_2(\ell)_{i, \mathrm{data}}$ is the 2REE of the $i$-th eigenstate and
$S_2(\ell)_{i, \mathrm{fit}}$ is a fitted value of it,
for all eigenstates of the non-integrable
Hamiltonian~\eqref{eigenstate_model} with $\Delta=2$ and $J_2=4$ (we consider only
the sector of a vanishing total momentum and magnetization). 
The eigenstates are sorted in descending order in terms of the residuals, 
and the horizontal axis represents their percentiles. 
The fits become better as the size of the system increases. 
{\bf d,} Same as figure {\bf c} for the integrable Hamiltonian ($\Delta=2, J_2=0$).
The fits become worse as the size of the system increases. 
}
\label{FigEigenstate}   
\end{figure*}

Figures~\ref{FigEigenstate}a,~\ref{FigEigenstate}b show the 2RPC
of the eigenstates of this model with various energies, which
are obtained by exact diagonalization.
We see that the fit by the formula~\eqref{FitFunc} works quite well for the non-integrable cases, although not for the integrable cases. 
Moreover, as Figs.~\ref{FigEigenstate}c,~\ref{FigEigenstate}d clearly indicate,
the residuals of the fits per site for all eigenstates decrease with respect to $L$ for the non-integrable cases
but increase for the integrable cases. We therefore numerically conclude that our formula~\eqref{FitFunc} is
applicable to non-integrable models but not to integrable models. 
We provide a discussion of the physics behind this result in Supplementary Note~2. 

We comment on our results of the energy eigenstates from the viewpoint of ETH.
First, the success of our formula in the non-integrable case is important for the following reasons: 
the corresponding thermal ``ensemble" is not the usual microcanonical ensemble
(mixed state)  but is rather 
a thermal pure state. 
Although the Page curve necessarily deviates from the volume-law slope, the way how it deviates always exhibits a universal behavior.
Furthermore, $S_2(\ell)$ is a highly non-local and complicated observable.
We thus expect that our results will bring the studies of ETH to the next step, i.e.,~its non-local extension. 
Second, with regards to the extension of ETH to non-local quantities, there are new proposals in which the effect of the energy fluctuation is incorporated \cite{Grover2015, Anatoly2016}, which is called subsystem ETH \cite{Anatoly2016}. 
In the subsystem ETH, the authors suggest that the volume-law slopes of the higher order REE for the energy eigenstates may be different 
from those of the Gibbs ensembles or the cTPQ state. 
We provide a discussion on this deviation in Supplementary Note~2.
In our numerical calculations on the energy eigenstates, however, we do not see the deviation of the 2RPC from \eqref{FitFunc}, which is derived for the cTPQ states.
This might be because of the limitation of the system sizes, and it would be interesting to see the deviation indeed occurs in larger systems.  
Indeed, this extension was recently analyzed in Ref.~\cite{TChengTarun}.
The EE of energy eigenstates in a non-integrable model was studied there
by substituting $\rho_{\rm diag}$ in Supplementary Eq.~(11) by the microcanonical density matrix (ensemble),
and the result supports our generalized formula, Supplementary Eq.~(11).
Third, the failure of our formula in the integrable cases is
surprising because ETH for local observables was proved to hold
for almost all eigenstates, even in integrable systems~\cite{weakETH_integrable}.
By contrast, our results in Figs.~\ref{FigEigenstate}c,~\ref{FigEigenstate}d clearly show
that almost all eigenstates actually violate formula~\eqref{FitFunc}, or non-local ETH.
A similar observation was made in Ref.~\cite{Vidmar2017},
where the EE of eigenstates in a different integrable model from ours was explicitly calculated.
They found that the subsystem size dependence of the EE is completely different from the result by Page, where the EE is close to its maximum value~\cite{Page:1993df}.
Their result is consistent with our finding that the 2REE of an integrable model is qualitatively different from that of a non-integrable model.

\subsection*{Numerical results for quenched states}
\begin{figure*}
   \includegraphics[width = 1.0\linewidth]{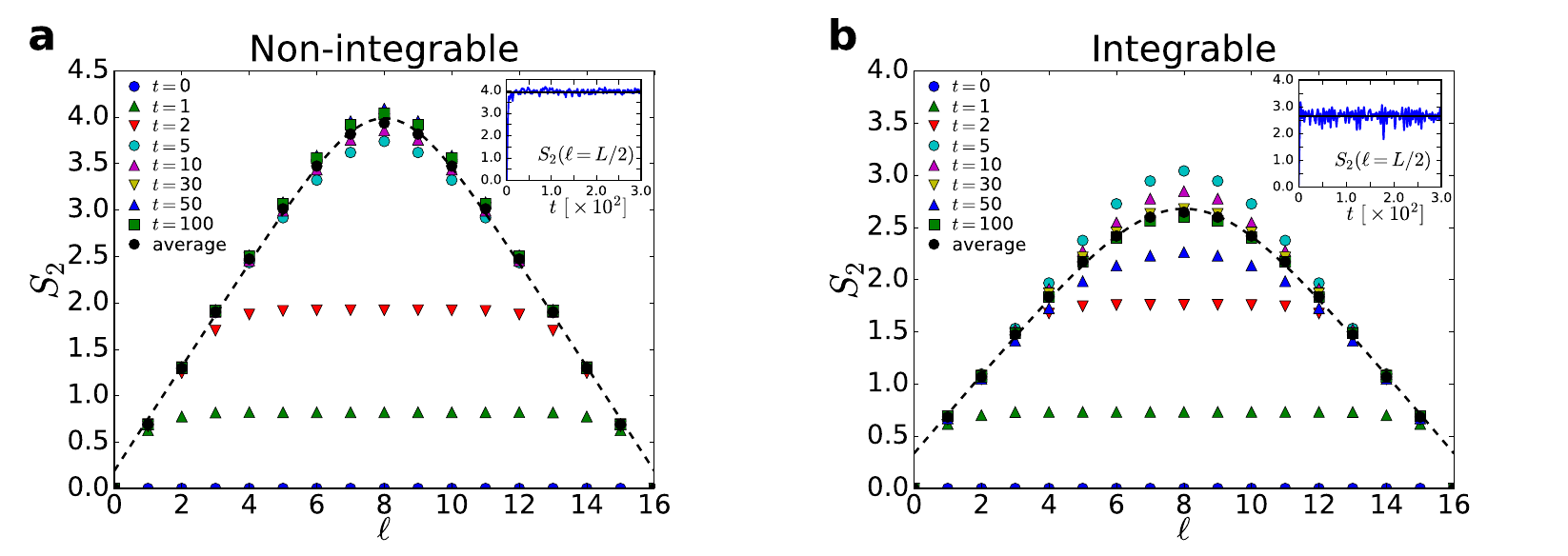}
\caption{ \label{Fig_quench}
{\bf  Dynamics of the second R\'{e}nyi Page curve after quantum quench.}
{\bf a,}
Time evolution of the 2RPC in a non-integrable system
(Eq.~\eqref{eigenstate_model} with $\Delta=1$ and $J_2=0.5$) after a quantum quench from the N\'{e}el state.
The dotted line is the fitting by Eq.~(\ref{FitFunc}) for the time average of $S_2(\ell)$.
The inset shows the dynamics of the 2REE at the center of the system, $S_2(L/2)$.
{\bf b,} Same as figure {\bf a} for the integrable Hamiltonian ($\Delta=1, J_2=0$). 
Eq.~(\ref{FitFunc}) fits the time average well in both {\bf a} and {\bf b}.
}   
\end{figure*}

Next, as a second example of thermal quantum states where our formula~\eqref{FitFunc} applies,
we consider pure states after quantum quenches in closed quantum systems, where 
some parameters of their Hamiltonians are abruptly changed~\cite{PolkovnikovReview2011, Eisert_Review2015, Calabrese_quench}.
When such states become stationary after a long time,  
they are considered to 
represent a thermal equilibrium corresponding to the Gibbs ensemble in non-integrable systems, 
although they do not in integrable systems
because infinitely many conserved quantities block thermal behaviors~\cite{GGE2007,Biroli2010}.
 
Here, we numerically simulate the dynamics of the 2RPC
after a quantum quench from pure states and
find that our formula~\eqref{FitFunc} explains well the 2RPC of the stationary states.
We note that experimental measurement of the dynamics of the 2RPC was already realized in Ref.~\cite{Science2016}. 
Indeed, our results explains the experimental data well: Fig.~4A in Ref.~\cite{Science2016} is well-fitted by our formula~\eqref{FitFunc}, and the parameters are estimated to be $\ln(a)=0.974$ and $ \ln K= 0.162$
(see Supplementary Figure~2).

We consider the Hamiltonian~(\ref{eigenstate_model}) with $\Delta = 1$, i.e., the Heisenberg model.
Again, we consider both integrable ($J_2=0$) and non-integrable ($J_2 = 0.5$) cases.
An initial state of a quantum quench ($t=0$) is taken as the N\'{e}el state
$\ket{{\rm N\acute{e}el}} \equiv \ket{\uparrow \downarrow \uparrow \downarrow \uparrow \downarrow ...}$, 
and the dynamics after the quench,
$\ket{\psi(t)} = e^{-iHt} \ket{\mathrm{N\acute{e}el}}$, is numerically calculated
by exact diagonalization.
We take a time step of the evolution as $dt = 0.1$ and calculate the dynamics of 
2RPC $S_2(\ell,t)$ up to $ t \leq T=300$.

In Fig.~\ref{Fig_quench}, we show the numerical results of the 2RPC $S_2(\ell,t)$ after the quench (see also  Supplementary Movies~1 and 2).
At first, the 2REE increases linearly with time until it saturates~\cite{Calabrese_quench, Eisert_Review2015,HYKim2013,Fagotti,Alba}. 
After it saturates, temporal oscillations of the 2RPC
are observed for a long time as long as $t = T =300$ (see the inset).
We consider these oscillations as being due to the finite size effect $(L=16)$, 
and thus, we also present the time average of the Page curve,
$\bar{S}_2(\ell) := \frac{1}{T} \int_0^T dt  S_2(\ell,t)$,
as an estimation of the long-time limit,  $\lim_{t\to\infty} S_2(\ell,t)$.
As clearly seen in Fig.~\ref{Fig_quench}, the time average of $S_2(\ell)$ is well fitted
by our formula~\eqref{FitFunc} (dotted line) for both non-integrable and integrable cases. 
This illustrates the validity of our formula for 
pure states after a quantum quench.

Although our formula~\eqref{FitFunc} is derived from a thermal state,  
it somehow works in the integrable cases where the states never thermalize. 
This success is achieved because the pure quantum states are scrambled and partially thermalized, which usually leads the states to the generalized Gibbs ensemble (GGE)~\cite{GGE2007}. 
Hence, the slope $\ln a$ quantifies the number of states in the GGE. 
We give a (non-rigorous) proof of the validity of the formula for the time-averaged 2RPC 
in Supplementary Note~2.
We also note several differences between the integrable cases and the non-integrable cases. 
First, for the integrable cases, other physical observables, such as staggered magnetization, 
are not explained by the cTPQ states or thermal equilibrium states~\cite{Pozsgay2013}.
Second, the properties of the temporal fluctuations are different between the two cases.
A fluctuation is larger in integrable cases than in non-integrable cases,
as one can see in the inset of Fig.~\ref{Fig_quench}, where the dynamics of
the 2RPC 
at the center of the system, $S_2(L/2)$, is plotted.
We observe that the fluctuation decays algebraically with the system size $L$ 
for the integrable cases, whereas it decays exponentially in the non-integrable cases~\cite{Kiendl2016}.
These differences probably reflect the existence of infinitely many conserved quantities
in integrable systems.

We also comment on the implication of our results
on recent studies on chaos in quantum many-body systems.
In Refs.~\cite{Ruihua2016,OTOCexpr}, the time dependence of $S_2(t)$ is related to
the out-of-time-ordered correlation, which
captures the essence of quantum chaos~\cite{Maldacena2016}.
As we discuss in Supplementary Note~2, 
the convergence of $S_2(\ell,t)$ in time and the applicability of formula~(\ref{FitFunc}) thereof
mean that the quantum states become scrambled in whole spatial regions. 
Therefore, we expect that the time until the convergence of the 2RPC after the quench
is a good diagnostic for the scrambling time~\cite{Lashkari2013}, and it would be interesting to
relate those two time scales directly in future work.

\subsection*{Application to many-body localization transition}
\begin{figure*}
   \includegraphics[width = 1.0\linewidth]{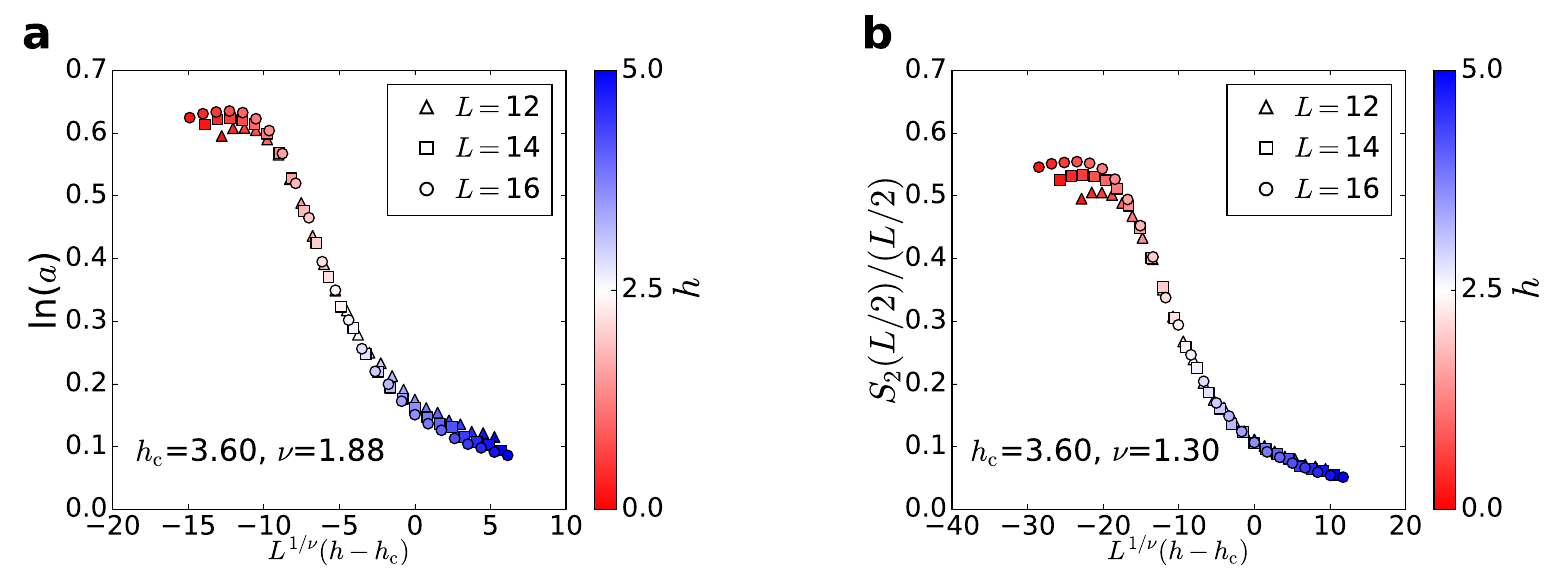}
\caption{ \label{Fig_MBL}
{\bf  Finite-size scaling across the ETH-MBL phase transition.}
{\bf a,} Finite-size scaling of $\ln(a)$, extracted from the fitting of the 2RPC of
the eigenstates of the Hamiltonian~\eqref{mbl_model}, versus $L^{1/\nu} (h-h_c)$. 
The estimation of the critical exponent $\nu$ is significantly improved.
{\bf b,} Same finite-size scaling as {\bf a} for
$s_{2, \mathrm{center}} = S_2(L/2)/(L/2)$, which is
a conventional estimate of the 2REE per site~\cite{MBLTransition2015}.
}
\end{figure*}
Thus far, we have shown that broad classes of thermal quantum states obey
our analytical formula~\eqref{FitFunc}.
Finally, we demonstrate its practical applications and advantages.
The example we take here is a problem of the phase transition between an ETH phase
and a MBL~\cite{MBLETH2016,MBLPalHuse2010} phase, where the volume law of entanglement provides an important diagnostic to
distinguish the two phases~\cite{Pollmann2014, MBLTransition2015}.

MBL is defined as a disorder-induced localization in interacting systems and
has recently been an active field of research because of the nontrivial 
interplay between the interactions and the disorder.
Meanwhile, the ETH phase is the phase where ETH holds, and it appears when
the strength of the disorder in a system is sufficiently weak (or absent).
There is considered to be a continuous phase transition between the ETH phase and the MBL phase
when the strength of disorder varies at a fixed energy density (temperature) of the system.  

Since MBL essentially involves interactions, it is difficult to study them analytically, 
and one often resorts to numerical approaches.
One of the simple diagnostics for ETH and MBL employed in such numerical studies
is the $n$-th R\'enyi EE per site $s_n$,
which will exhibit a transition from being non-zero (ETH phase) to being zero (MBL phase) in the thermodynamic
limit. Most previous studies have utilized $s_{n, \mathrm{center}} = S_n(L/2)/(L/2)$~\cite{Pollmann2014, MBLTransition2015,PollmannTiEv2012, Singh2016},
the EE per site at the center of the system, as an estimate of $s_n$.
As we see in the inset of Fig.~\ref{Fig TPQ}, however, the $s_{n, \mathrm{center}}$ includes a systematic error of $O(1/L)$ 
from the correct value of $s_n$ in the thermodynamic limit
because of the deviation of the $n$RPC from the volume law at the center of the system.
This error harms analyses of the ETH-MBL transition since the system size $L$ accessible by numerical methods
is not very large, $L \sim 22$. 

Here, our formula~\eqref{FitFunc} comes in, and extracts the 2REE per site in the thermodynamic
limit in relatively small systems, as we illustrated in Fig.~\ref{Fig TPQ}
(we give a brief justification of the validity of our formula in disordered systems in Supplementary Note~3).
To substantiate the advantage of our formula~\eqref{FitFunc},
we study the ETH-MBL transition in the prototypical $S=1/2$ spin chain for MBL~\cite{MBLPalHuse2010, MBLBauerNayak2013,MBLTransition2015}, 
\begin{align}
H = \sum_{i=1}^L \left( \bm{S}_i \cdot \bm{S}_{i+1}+ h^z_i S^z_i \right),
\label{mbl_model}
\end{align}
where a random magnetic field $\{ h^z_i \}_i$ is drawn from a uniform distribution [$-h$, $h$] and a periodic boundary condition is imposed. 

This model exhibits the ETH-MBL phase transition at the critical disorder strength $h_\mrc$.
Using the von~Neumann EE ($S_{n=1}$), the authors of Ref.~\cite{MBLTransition2015} estimated 
$h_\mrc=3.62 \pm 0.2$ and the critical exponent of the transition $\nu = 0.80 \pm 0.4$ at the center of the energy spectrum.
However, the estimated $\nu$ violates the Harris bound $\nu \geq 2$~\cite{MBLBounds2015}.
This violation can probably be understood as a finite-size effect,
and we will show that the usage of formula~\eqref{FitFunc} improves the situation.
We note that the Harris bound for $S_2$ is the same as that of $S_{n=1}$~\cite{MBLBounds2015}.

In Fig.~\ref{Fig_MBL}a, we perform a finite-size scaling of the 2REE per site, $s_2$, for the first time. 
Those data are extracted from the fitting of the 2RPC $S_2(\ell)$ of the eigenstates of the system, namely, $s_2 = \ln(a)$ 
at the center of the energy spectrum. 
We obtain $h_\mrc=3.60 \pm 0.12$ and $\nu=1.88 \pm 0.2$ (details are presented in the Method section). 
Although we study smaller systems, up to $L=16$, compared
with the previous study~\cite{MBLTransition2015}~($L \sim 22$),
the estimation of $\nu$ exhibits a significant improvement to satisfy the Harris bound $\nu \geq 2$.
This 
highlights the usefulness of our formula~\eqref{FitFunc}
since a finite-size scaling of the conventional method, $s_{2, \mathrm{center}} = S_2(L/2)/(L/2)$, under the same conditions
yields $h_\mrc = 3.60 \pm 0.12$ and $\nu = 1.30 \pm 0.12$ (Fig.~\ref{Fig_MBL}b). 

In general, when a physical quantity in a finite-size system has an $O(1/L)$ difference from the quantity in the thermodynamic limit,
this difference deteriorates the finite-size scaling, and the critical exponent estimated from the scaling becomes completely different from that in the thermodynamic limit (for example, see the conventional ferromagnetic-spin glass transition~\cite{FSS2006, FSSHukushima}).
In this sense, the critical exponent $\nu$ in our study is also sensitive to the finite-size effect of $O(1/L)$.  
However, using formula~\eqref{FitFunc},
we can remove the finite-size effect of $O(1/L)$, as shown in Fig.~\ref{Fig TPQ}, resulting in an improvement in the estimation of $\nu$. 

\section*{Discussion}
In conclusion, we studied the volume law of entanglement in general scrambled pure quantum states.
By employing the cTPQ states, we derive analytical formulae for the $n$RPC for general local Hamiltonians at any temperature. In particular, we show that the 2RPC $S_2(\ell)$ is parametrized by only two parameters (Eq.~\eqref{FitFunc}), and our formula improves the finite-size scaling of the thermodynamic quantities.
We numerically demonstrate that the same formula for the 2RPC works as a fitting function for general thermal quantum states
other than cTPQ states, namely, excited eigenstates of general Hamiltonians and states after quantum quenches.
We also propose the generalized formula which can incorporate the energy fluctuation in Supplementary Eq.~(11).
Finally we employ our formula to detect the ETH-MBL phase transition. The full characterization of the 2RPC by our formula improves the estimate of the critical exponent of the transition and would resolve the controversy over the breakdown of the Harris bound at the transition.

\section*{Methods}
\subsection*{Derivation of Equation~\eqref{FitFunc}}
Here we present the detailed calculation of $n$-th R\'{e}nyi entanglement entropy (REE) of
the canonical thermal pure quantum (cTPQ) states~\cite{Sugiura2012, Sugiura2013}.

For any local observable $\mathcal{O}$ on $\mathcal{H}_\Sigma$, its expectation value of a cTPQ state satisfies
$\overline{ \braket{\psi | \mathcal{O}| \psi} } = \Tr(\mathcal{O} e^{-\beta H}) / \Tr(e^{-\beta H})$,
and the standard deviation from the average is exponentially small with respect to the volume of the system~\cite{Sugiura2013}
(here we denote a random average over the coefficients $\{z_j\}$ by $\overline{\cdots}$). 
In this sense, we can regard cTPQ states as a canonical example of thermal pure states which corresponds to the thermal Gibbs ensemble  $\rho_\mathrm{Gibbs} \equiv e^{-\beta H} / \Tr(e^{-\beta H})$ at inverse temperature $\beta$.

The reduced density matrix of the cTPQ state $\ket{\psi}$ in A is written as
\begin{equation}
    \rho_\mrA =  \trB \ket{\psi} \bra{\psi}
    = \frac{1}{Z_\psi}  \sum_{a_1,a_2, b_1, i_1, j_1} \pi_{12} \ket{a_1} \bra{a_2},
     \label{rho A}
\end{equation}
where
$\pi_{pq} \equiv z_{i_p} z_{j_p}^* \braket{ a_p b_p |	e^{-{1\over 2} \beta H} | i_p }
	\braket{ j_p | e^{-{1\over 2} \beta H} | a_q b_p }$,
$\ket{ab} \equiv \ket{a} \otimes \ket{b}$, 
and $\{\ket{a} \}_a$ and $\{\ket{b} \}_b$ are complete orthonormal bases in the subsystem A and B, respectively.
The two indices $i_p, j_q \in \{\Ket{j} \}_j$ run over the orthonormal basis $\{\Ket{j} \}_j$.
In this notation we obtain
\begin{align}
	&\trA\rho_\mrA^n
	=\frac{\sum_{(a),(b),(i),(j)}
		\pi_{12} \pi_{23} \cdots \pi_{n1}}
	{Z_\psi ^n},
\label{trcon}
\end{align}
where $(x) \equiv x_1,x_2, \cdots, x_n$.
What we want to calculate is a random average of  $n$-th REE, 
$\overline{S_n} := \dfrac{1}{1-n} \overline{ \ln \left( \trA  \rho_\mrA^n \right) }$.
The term $\pi_{12} \pi_{23} \cdots \pi_{n1} $, however, includes the product of random variables such as 
$z_{i_1} z_{j_1}^* z_{i_2} z_{ j_2}^* \cdots z_{i_n} z_{ j_n}^*$ and 
it is difficult to calculate the average of the logarithm of it.
Instead, we calculate $S_n$ by averaging the trace before taking the logarithm of it,
$\tilde{S}_n := \dfrac{1}{1-n}  \ln \left( \overline{ \trA  \rho_\mrA^n \ } \right)$.
As we give a proof in Supplementary Note~4, the difference between $\overline{S}_n$ and $\tilde{S}_n$ is
exponentially small in terms of the system size $L$ (see also Ref.~\cite{TChengTarun}).

By taking the random number average of $\trA  \rho_A^n $ with using several properties of $\{z_i\}$ such as
$\overline{z_i} = 0, \: \overline{z_i^* z_j} = \delta_{ij}$ and $\overline{|z_i|^2 |z_j|^2} = 1 + \delta_{ij}$,
we can calculate $n$-th REE of the cTPQ state for any Hamiltonian (below we just use $S_n$ to denote $\tilde{S}_n$). 
For example, the 2REE is Eq.~\eqref{TPQ_S2}, and 
the third REE is 
\begin{align}
  S_3  = -\frac{1}{2} \ln  \left[  \left( 
     \trA{ \left(  \trB e^{-\beta H} \right)^3 }  + 3\Tr M  \right. \right. \nonumber\\
   \left. \left.
     + \trB{ \left( \trA e^{-\beta H} \right)^3 } + N \right) \bigg/ (\Tr{e^{-\beta H}})^3
	\right],
%
\end{align}
where $M=
e^{-\beta H}\left(\trB e^{-\beta H} \otimes\trA e^{-\beta H} \right)$
and $N=\sum_{} 
\braket{a_1 b_1 | e^{-\beta H}| a_3 b_2}
\braket{a_2 b_2 | e^{-\beta H}| a_1 b_3}
\braket{a_3 b_3 | e^{-\beta H}| a_2 b_1}
$. 
After the same simplification as we do for the second REE, the term $N$ is shown to be exponentially smaller than the other terms. Then, we obtain the final result Eq.~\eqref{S3_simple}.
It is difficult to write down a general expression of $S_n$ for all $n$, but
we can calculate it for given integer $n \geq2$ systematically by taking an average in Eq.~\eqref{trcon}.
It would be interesting to study whether the universality which we reported for the 2REE
holds for the entanglement entropy ($n=1$).

Finally, we also comment on the case of infinite temperature $\beta=0$ and its relation
to the previous studies.
When $\beta=0$, we can obtain a fairly simple equation for general $n$:
\begin{equation}
	S_n=\ell\ln{2}-\frac{1}{n-1}\ln\left[{\sum_{k=1}^{n} N(n,k)\left(\frac{2^\ell}	
	{2^{L-\ell}}\right)^{k-1}}\right],
\end{equation}
where $N(n,k)=\frac{1}{n}\binom {n}{k} \binom {n}{k-1}$ is known as the Narayana numbers.
It is possible to take an analytic continuation $n\to 1$ and reproduce the result
on the entanglement entropy $S_1 = - \trA \left( \rho_\mrA \ln \rho_\mrA \right)$ calculated in Ref.~\cite{Page:1993df}.

\subsection{Numerical fitting by our formula~\eqref{FitFunc}}
Throughout this work, numerical fitting by our formula~\eqref{FitFunc}
for given data of the 2RPC $\{ S_2(\ell) \}_{\ell=0}^{L}$
is performed with the least squares method implemented in the numerical package
\verb|scipy.optimize.leastsq|  by regarding $a$ and $K$ in Eq.~\eqref{FitFunc}
as fitting parameters.

\subsection{Details on the numerical calculations on ETH-MBL transition}
In the demonstration of the ETH-MBL transition, 
we did exact diagonalization on the Hamiltonian~\eqref{mbl_model}
$ H_\mathrm{MBL} = \sum_{i=1}^L \left( \bm{S}_i \cdot \bm{S}_{i+1}+ h^z_i S^z_i \right) $,
where random magnetic fields $\{ h_i^z\}_i$ are taken from a continuous uniform distribution within $[-h, h]$
and periodic boundary condition is imposed.
Since the Hamiltonian conserves the total magnetization $S^z_\mathrm{tot} = \sum_i S_i^z$,
we fixed the magnetization sector as $S^z_\mathrm{tot} = 0$.
We define energy density $\epsilon$ of the eigenstates with energy $E$
by using the extremal values of energies,
$\epsilon = (E - E_\mathrm{min})/(E_\mathrm{max} - E_\mathrm{min})$,
and focus on the case of $\epsilon =0.5$ throughout our analysis.
For each realization of the magnetic fields, an energy eigenstate closest to $\epsilon=0.5$
is chosen and the 2RPC $S_2(\ell)$ for that eigenstate is calculated.
The number of realizations of the random magnetic fields is 1000 for all presented data.
After averaging $S_2(\ell)$ over all realizations, we perform fitting of $S_2(\ell)$ by our formula~\eqref{FitFunc}
and obtain $\ln(a)$ for each value of the disorder strength $h$.

The scaling analysis of $\ln(a)$ and $s_{2, \mathrm{center}} = S_2(L/2)/(L/2)$ is 
performed to determine the critical disorder strength $h_c$ and the critical exponent $\nu$
in a following procedure~\cite{MBLTransition2015}.
First we assume a universal scaling function of the form $g( L^{1/\nu}(h-h_\mrc))$
in a window of width $2w$ centered at $h_\mrc$, namely, $[h_\mrc - w, h_\mrc + w]$. 
The function $g$ is approximated by a polynomial of degree three and
the polynomial as well as $h_\mrc$ and $\nu$ are optimized to make all data collapse into one line.
In our analysis $w=2.0$ is used,
and $(h_\mrc, \nu) =(3.60 \pm 0.12, 1.88 \pm 0.2)$ is found for the case of $\ln(a)$
while $(h_\mrc, \nu) =(3.60 \pm 0.12, 1.30 \pm 0.13)$ for $s_{2, \mathrm{center}}$
(the errors are estimated by a bootstrap method).

\subsection*{Data availability}
All numerical data and computer codes used in this study are available
from the corresponding author upon request.


\section*{Author contributions}
The analytical calculations were mainly performed by MW and SS.
The numerical calculations were mostly performed by YON with the support of HF and SS.
SS conceived and supervised the project. 
All authors extensively discussed the results and prepared the manuscript.

\section*{Competing financial interests}
The authors declare no competing financial or non-financial interests.

\section*{Acknowledgement}
The authors thank J.~Eisert, F.~Pollmann, T.~Grover, M.~Oshikawa, T.~Sagawa, T.~Takayanagi, and H.~Tasaki for their valuable discussions. 
The authors thank T.~N.~Ikeda and K.~Kawaguchi for their in-depth readings of the manuscript. 
The authors gratefully acknowledge the hospitality of the Yukawa Institute for Theoretical Physics at Kyoto University,
where this work was initiated during the long-term workshop YITP-T-16-01
``Quantum Information in String Theory and Many-body Systems''.
HF, YON and SS also gratefully acknowledge the hospitality of the Kavli Institute for Theoretical Physics,
where this work was improved and supported in part by the National Science Foundation under Grant No. NSF PHY-1125915.
The authors are supported by JSPS KAKENHI Grant Numbers
JP16J04752, JP16J01135, JP15J11250 and JP16J01143, 
and by World Premier International Research Center Initiative (WPI), MEXT, Japan.
HF and YON also acknowledge support from the ALPS program
and MW from the FMSP program, both under ``The Program for Leading Graduate Schools'' of JSPS.

%
\end{document}